\documentclass{article}
\input epsf.sty	%figures program compatible with Apple and Sun
\usepackage{amsfonts} % graphics 

\date  {{\small Oct. 21, 1997, updated March 20, 1998.}} 

\title{Continuum Limits for Critical Percolation  \\    
and Other Stochastic Geometric Models}

\author{M. Aizenman  
\footnote{Permanent address: 
  Departments of Physics and Mathematics, Jadwin Hall, 
  Princeton University, Princeton, NJ 08544-0708, USA.
            E-mail:  aizenman@princeton.edu} 
  } 

\begin{document}
 
 % heading:   
    \begin{figure}[t]{ \footnotesize   
    For: Proceedings Int. Cong. Math. Phys. (Brisbane, 1997)}
    \end{figure} 

 % footnote (no number)
  \renewcommand{\baselinestretch}{1}
  {
  \renewcommand{\thefootnote}{}
  \footnotetext{Report prepared at the Centre {\'E}mile Borel, 
  Institute Henri Poincar\'e, Paris.}
            }
  
  % OK - back on track  
  
\maketitle

\abstract{The talk presented at ICMP 97 focused on the 
scaling limits of critical percolation models, and some other  
systems  whose salient features can be described by collections  
of random lines.  In the scaling limit we keep track of
features seen on the macroscopic scale, in situations where the 
short--distance scale at which the 
system's basic variables are defined is taken to zero. 
Among the challenging questions are the construction of the limit, 
and the explanation of some of the emergent properties, 
in particular the behavior under conformal maps as discussed in 
ref.~\cite{LPS}.  
A descriptive account of the project, and some related open 
problems, is found in ref.~\cite{Aiz_IMA} and in 
\cite{AizBurch} (joint work with A. Burchard) where tools are 
developed for establishing a curve--regularity condition 
which plays a key role in the construction of the limit. 
The formulation of the scaling limit as a 
random Web measure permits to formulate the question of uniqueness 
of measure(s)  describing systems of random curves satisfying 
the conditions of independence, Euclidean invariance, and 
regularity.  The {\em uniqueness} question remains open; 
progress on it could shed light on the purported 
universality of critical behavior and the apparent 
conformal invariance of the critical measures.  
The random Web yields also another perspective on some of the 
equations of conformal field theory which have appeared in this 
context, such as the equation proposed by J. Cardy \cite{Car}.  }

\section{Comments on Selected Points}
 
The connected clusters, conveniently viewed as the 
conducting clusters, in 
{\em critical} percolation models are well known to assume 
``fractal'' characteristics  \cite{Man}, extending 
tenuously over many scales.  
Our goal is to formulate a  macroscopic description of 
such systems in terms which remains meaningful 
in the scaling limit, where the scale at which the model 
is constructed drops out of sight (is taken to zero).  
The interest in this question is enhanced by the 
observation that the correlations in various spin models   
have a geometric representation in terms of random cluster 
with somewhat similar characteristics \cite{FK,AizNa}, which 
suggests that the resulting system of random collections of 
lines may bear  interesting relations to certain random fields.  
There is accumulating evidence for 
conformal invariance of the critical models 
\cite{LPS,Car,Pin,Wat}, and specific
equations were surmised on the basis of analogies
with random fields.
It would be interesting to understand such equations 
in terms which are native to the microscopic model.  

Following are few words to indicate some of the relevant
points.  Further discussion, and a better reference list 
is provided in Ref.~\cite{Aiz_IMA}, from which 
also the figures are taken,  and Ref.~\cite{AizBurch}.

\noindent $\bullet$ \qquad There are differences between 
the {\em \bf  macroscopic} 
and the {\em \bf microscopic} perspectives on a physical 
system.  While the difference is not hard to understand, 
ignoring it one runs the peril of paradoxes 
(ranging from Zeno's puzzles up to some more recent 
lively discussions concerning the non-uniqueness of the 
Incipient Spanning Cluster(s)~\cite{Aiz.ISC}.   

% Figure 1 
\begin{figure}[htb]
    \begin{center}
    \leavevmode
    \hbox{%
    \epsfxsize=3.8in 
    \epsffile{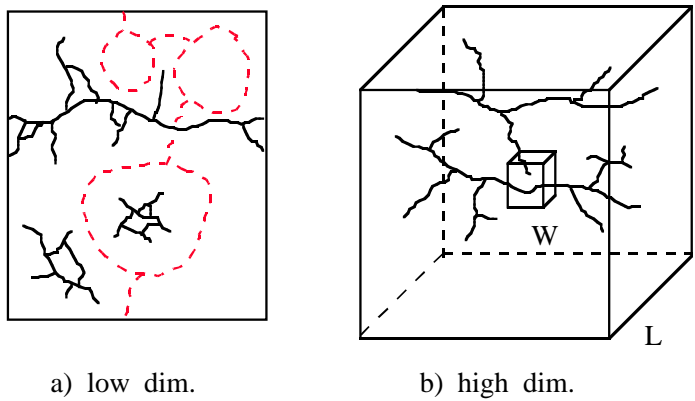}  }
\end{center}  
\caption{(After ref.~\cite{Aiz_IMA})
   Schematic depiction of the macroscopic view of the 
   connected clusters in critical percolation models.  a) In $d=2$, 
   and presumably other ``low dimensions'' on each scale one would  
   typically find a finite number of distinct clusters  
   within a given compact region.  The obstructions are likewise 
   present.  
   b) In high dimensions (presumably all $d>6$) the clusters 
   proliferate and extend: the probability that there are at least 
   $k$ spanning cluster tends to one for each $k < \infty$, 
   even if one considers only clusters touching a preassigned region 
   whose size shrinks  as $W\approx \delta^{|d-6|_{+}/(d-4) - o(1)}$
   or alternatively cluster reaching exceeding distances away.}
\label{fig:exit}
\end{figure}

The microscopic view has been mathematically formulated, 
and extensively studied, within the  infinite-system formalism.  
New constructs are required for the macroscopic view.  
Field theory provides the relevant tool for systems with order 
parameter. However, there is room and need for other tools
to capture the 
stochastic geometric features seen, e.g., in percolation models.

\noindent $\diamond$ \qquad  
The seemingly natural formulation, in which the geometric features 
are expressed through  the random clusters, 
viewed as closed subsets of the Euclidean space 
with the notion of distance based on the Hausdorff metric, 
is inadequate.  
The problem is caused by the ubiquity of choke points: 
typical configurations exhibit many sites at which a 
microscopic change (often just the flip of a single bond) 
produces a macroscopic scale effect.  

\noindent $\bullet$ \qquad 
It is proposed to base the macroscopic description of such 
models on the  collection, henceforth called the {\em Web}, 
of  all the realized (connecting) curves, which 
are self-avoiding in the suitable sense and
supported on the connected clusters.  
Alternatively, though this notion is less developed at 
present,  one may describe the configuration in terms 
of the connected clusters but change the notion of distance 
from the Hausdorff metric to one based on the connected 
curves they support. For convenience, in 2D
one may simultaneously keep track of the dual system of 
obstructions, which forms another family of curves, with 
the natural non-crossing condition between the two curve systems.

\noindent $\diamond$ \qquad 
Some degree of regularity (uniform in the short scale distance
$\delta$) is needed in order for a description based on random 
curves to remain meaningful in the scaling limit ($\delta \to 0$).
A generally applicable criterion is developed in ref.~\cite{AizBurch}, 
where the following result is shown to be implied by a 
condition, defining Type 1 critical models, which is known to 
hold for the $2D$ percolation models.  
(That the $2D$ models are of {\em Type 1} is the result of the 
Russo~\cite{R} and Seymour--Welsh~\cite{R,SW} theory,
which was recently adapted to a model with spherical--symmetry 
by K. Alexander~\cite{AlexRSW}.)   
%%%%%

\bigskip  % sorry -- needed here
\noindent {\bf Theorem 1} (Regularity of connecting curves)
{\it For the two-dimensional site, or bond, critical 
percolation model, on $\delta  {\bf Z}^{2}$ ($\delta < 1$), 
there is a H\"older continuity exponent,
$\alpha > 1/2$ $(= 1/d)$ such that all the self-avoiding 
paths (polygonal with step size $delta$]) along the connected 
clusters in  $[0,1]^{2}$ 
can be simultaneously parameterized by continuous functions
$\gamma (t)$, $0\le t\le 1$ , satisfying }
\begin{equation}
| \gamma (t_{1}) - \gamma (t_{2}) | \ 
 \le \kappa(\omega) \ |t_{1}-t_{2}|^{\alpha}; ,\quad
 \mbox{for all \ $0\le t_{1} < t_{2} \le 1$,}
\end{equation}
{\it where $\kappa$ is a random variable whose probability 
distribution does not drift to infinity as $\delta \to 0$: }
\begin{equation}
Prob_{\delta}\left( \kappa(\omega) \ge u \right) \  \le \ g(u)
\end{equation}
{\it with  
$g(u) {\parbox[t]{.4in} {$\longrightarrow\\[-9pt]   
  {\scriptstyle u\to \infty}$}} 0$  uniformly in $\delta$. 
  }        % end theorem 
\bigskip   % sorry -- here \bigskip is needed

This statement provides the required 
{\em tightness} of the probability distribution of the Web, 
which is essential for the existence of a limiting measure, 
in the scaling limit $\delta \to 0$. 
The regularity can alternatively be expressed in 
terms which are manifestly parametrization independent, 
as upper bounds on the curves' ``tortuosity''. 

\noindent $\bullet$ \qquad 
Lower bounds on the {\em tortuosity} of the curves are also of some 
interest, and some such bounds are established, at a similar degree 
of generality~(\cite{AizBurch}).

\noindent $\diamond$ \qquad 
The construction leads to an existence result: 

\bigskip
\noindent{\bf Theorem  2}  
{\it For $d=2$ dimensions, and more generally 
for each dimension in which the critical behavior is 
Type I, there is a one--parameter family of probability
measures ($\mu_t$) on the space of collections of curves 
(satisfying suitable consistency conditions)  
each of which has the following properties. \\
1. {\em (Independence)} For disjoint closed regions,  
$A,  B \subset R^d$, $W_A(\omega)$ and $W_B(\omega)$
(the restrictions of $W$ to curves in the indicated regions) 
are independent.     \\ 
2. {\em (Euclidean invariance)}  The probability measure is 
invariant under translations and rotations.   \\ 
3. {\em (Regularity)} 
The spanning probabilities of compact rectangular
regions are neither $0$ nor $1$:
\begin{equation}
R_{s}:= Prob{[\mu]}\left(
\begin{array}{c}
        \mbox{the web configuration $W$ includes a path }  \\
        \mbox{in $[-s,s]^d$ which crosses the cube left 
        $\leftrightarrow$ right}
\end{array}  \right) \ \ 
\stackrel{\textstyle > \; 0}{_{_{\textstyle < \; 1} }} 
\end{equation}  
 }  
\bigskip % end theorem

%  Figure 2
\begin{figure}[htb]
    \begin{center}
    \leavevmode
    \hbox{%
    \epsfxsize=2.5in 
    \epsffile{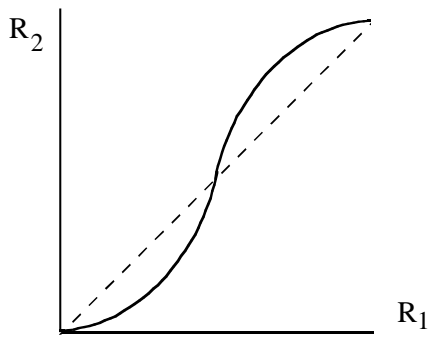}  }
\end{center}  
\caption{Schematic depiction of 
   the expected shape of the set of values attained jointly by 
   $R_1$ and $R_2$ within the one parameter family of random Web 
   measures.  The set is expected to form the graph of a map of 
   the interval $[0,1]$ onto itself with one unstable fixed point,
   corresponding to the critical case.  The slope at the unstable 
   fixed point ($=4/3 $?) yields a critical exponent. 
     }
\label{fig:RGmap}
\end{figure}

A convenient parametrization for these measures is 
the crossing  probability $R_{1}$. 
The measures are constructed as continuum 
limits of sequences 
of models with suitably adjusted percolation densities.  If the 
standard picture is correct, the density needs be adjusted with the 
lattice spacing as~\cite{BCKS}:
\begin{equation}
p_{\delta} \ = \ p_c + t   \ \delta^{1/\nu} \ .
\end{equation} 
In order to produce rotation invariant measures the construction 
is based on the droplet percolation model.  

\noindent $\bullet$  \qquad  
A direct approach to the continuum theory could be facilitated 
by a uniqueness statement, which may be presented as an open 
problem: does  the above family contain all the probability 
distributions of a random Web 
(the space needs to be defined more carefully than is possible 
here) having the properties 1) - 3).  
In general, we are short on arguments proving 
{\em uniqueness}.  Progress in that area could shed light on 
a number of issues.

\noindent $\diamond$  \qquad  
The standard renormalization-group picture would require the 
collection of the joint values of $R_1$ and $R_2$ attainable within 
the above family of Web measures to form a graph of the form 
depicted in Fig.~2.  This observation, which 
has not been established, shows that the one parameter family of 
measures described above may be viewed as representing the  
unstable manifold of the renormalization group flow 
at the fixed point corresponding to the critical model(s).  That of 
course should not be taken too literary since still is no 
space was identified in which this statement can be properly made.  

\noindent $\bullet $  \qquad  
There is  accumulating evidence for the conformal invariance of 
the special scale invariant measure in the above class.  
The proof can also be reduced to a strong enough uniqueness result 
\cite{BS,Aiz_prep}.  

 % Figure 3. 
\begin{figure}[htb]
    \begin{center}
    \leavevmode
    \hbox{%
    \epsfxsize=3.5in 
    \epsffile{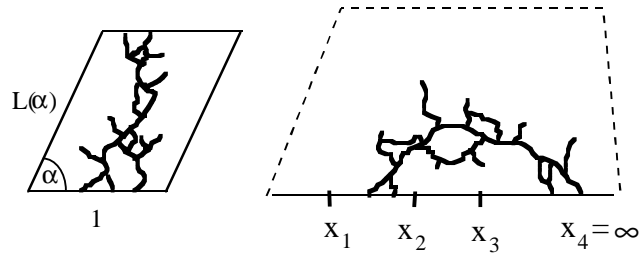}  }
\end{center}  
\caption{
The crossing event for whose probability 
an equation  was proposed by Cardy \cite{Car}, 
for the case $p=p_{c} \delta \to 0$.  
The two forms depicted here are related by 
a conformal map.  Among the goals of the present project is to 
define the random system whose probabilities are exactly 
such limits, and to explain equations like the proposed one
from within the microscopic model. }
\label{fig:LC}
\end{figure}

\noindent $\diamond$  \qquad  
Specific equations have been proposed, and tested, for 
the crossing probabilities of certain events, such as depicted
in Fig.~3, Ref.~\cite{Car,LPS,Pin,Wat}.  
These proposals are based on insights derived from other models 
with known relations to conformal field theory.  
It should be of interest to understand the equations from 
within the microscopic picture (for Ising spin systems 
such a project was started in \cite{KadCeva}).  Some ideas 
will be presented in \cite{Aiz_prep} (in preparation). 
This goal is somewhat reminiscent of the task of deriving 
the equations of hydrodynamics starting from the microscopic 
description of gases/fluids, a topic which 
was discussed in this conference by H-T Yau.

 % (Figure 4). 
\begin{figure}[htb]
    \begin{center}
    \leavevmode
    \hbox{%
    \epsfxsize=4in 
    \epsffile{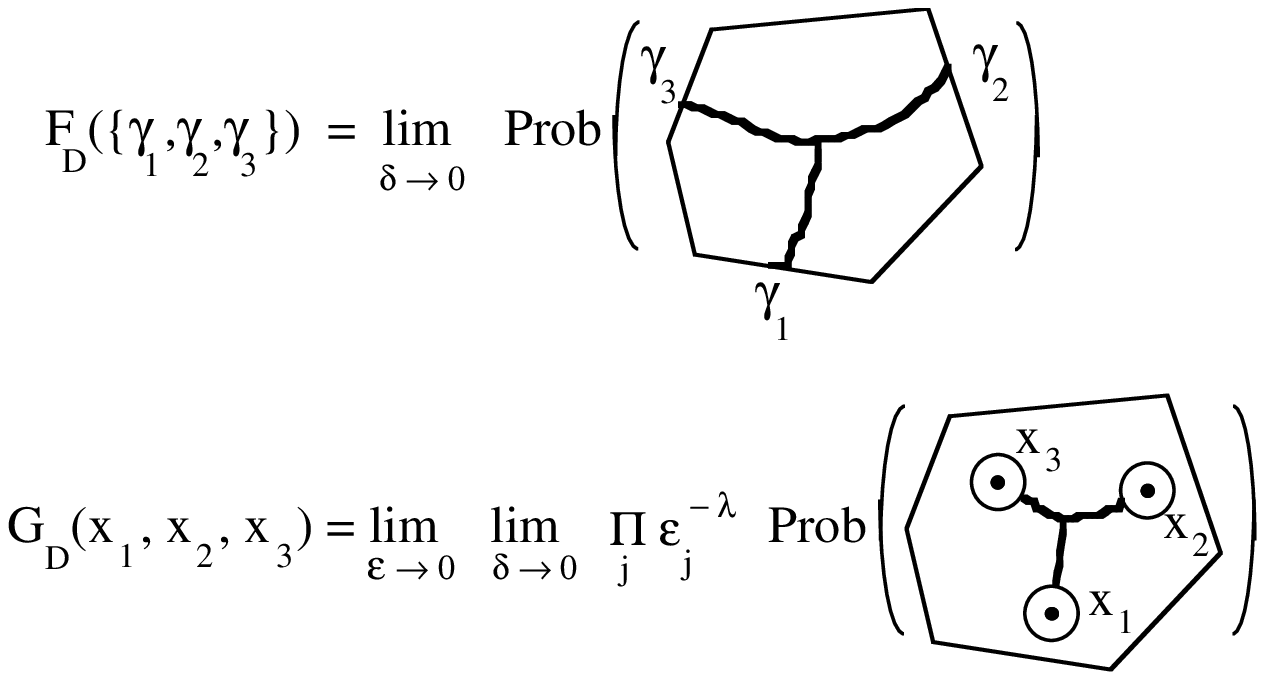}  }
\end{center}  
\caption{Two sets of functions which convey quantifiable 
information on the distribution of the random Web. }
\label{fig:FandG}
\end{figure}

\noindent $\bullet $  \qquad  
There are similarities and relation of the Web with some 
field theories.  The functions depicted in Figure~4 are 
somewhat reminiscent of vacuum expectation 
values of certain field operators (of different type for 
the two cases, the first one discussed in ref.~\cite{Car}).  
Intriguing possibilities are suggested by the 
Fortuin--Kasteleyn random cluster models, which 
cast Potts models in stochastic geometric terms, and 
interpolate between the Ising model, whose scaling limit 
yields the $\varphi_4$ field theory, and independent percolation
models whose scaling limit is the random Web.

\noindent $\diamond $  \qquad  
While the above discussion focused on percolation models 
where macroscopic fractal structures appear only when 
the density parameter is fine--tuned close to its critical value 
($p_c$),  such effects can appear also 
through  ``self-organized'' criticality \cite{BTW}.   
An examples to which some of the results described here 
are also  relevant is provided by the {\em minimal spanning tree} 
process~\cite{CCN,AlexMST,ABNW}.
% (work in progress with A. Burchard, C.M. Newman and D. Wilson). 

\section{Acknowledgments}
It is a pleasure to thank A. Burchard,  C.M. Newman, and D. Wilson 
for their contributions through collaborations in related works. 
I am also indebted to many colleagues for discussions 
of pertinent topics.  In particular I wish to thank here
R. Langlands, T. Spencer, A. Aharony, and D. Stauffer for 
stimulating comments.
The work was supported in part by the NSF Grant PHY-9512729. 

%\bibliography{percolation.Sept97}   \end{document}   

\begin{thebibliography}{10}

\bibitem{LPS}
R.~Langlands, P.~Pouiliot, and Y.~Saint-Aubin, ``Conformal invariance in
  two-dimensional percolation,'' {\em Bull. AMS},  {\bf 30},  1 (1994).

\bibitem{Aiz_IMA}
M.~Aizenman, ``Scaling limit for the incipient spanning clusters,'' in {\em
  Mathematics of Materials: Percolation and Composites} (K.~M. Golden, G.~R.
  Grimmett, R.~D. James, G.~W. Milton, and P.~N. Sen, eds.), The IMA Volumes in
  Mathematics and its Applications, Springer-Verlag (to appear).
\newblock [cond-mat/9611040] 

\bibitem{AizBurch}
M.~Aizenman and A.~Burchard, ``H\"older regularity and dimension bounds 
for random curves,'' 1998 preprint  [math.FA/9801027].

\bibitem{Car}
J.~Cardy, ``Critical percolation in finite geometries,'' {\em J. Phys. A},
  {\bf 25},  L201 (1992).

\bibitem{Man}
B.~Mandelbrot, ``Fractals in physics: Squig clusters, diffusions, fractal
  measures, and unicity of fractal dimensionality,'' {\em J. Stat. Phys.},
  {\bf 34},  895 (1984).

\bibitem{FK}
C.~Fortuin and P.~Kasteleyn, ``On the random cluster model. {I}. introduction
  and relation to other models,'' {\em Physica},  {\bf 57},  536 (1972).

\bibitem{AizNa}
M.~Aizenman and B.~Nachtergaele, ``Geometric aspects of quantum spin states,''
  {\em Commun. Math. Phys.},  {\bf 164},  17 (1994).

\bibitem{Pin}
H.~Pinson, ``Critical percolation on the {T}orus,'' {\em J. Stat. Phys.},  {\bf
  75},  1167 (1994).

\bibitem{Wat}
G.~M.~T. Watts, ``A crossing probability for critical percolation in two
  dimensions,'' {\em \em J. Phys. A},  {\bf 29},  L363 (1996).

\bibitem{Aiz.ISC} 
M.~Aizenman, ``On the number of incipient spanning clusters,'' {\em Nucl. Phys.
  B [FS]}, \nolinebreak[4] {\bf 485},  
   551  \nolinebreak[4] (1997). 
\bibitem{R}
L.~Russo, ``A note on percolation,'' {\em Zeit. Wahr.},  {\bf 43},  39 (1978).

\bibitem{SW}
P.~Seymour and D.~Welsh, ``Percolation probabilities on the square lattice,''
  in {\em {\em Advances in Graph Theory} Annals of Discrete Mathematics}
  (B.~Bollob{\'a}s, ed.), vol.~3, North Holland, 1978.

\bibitem{AlexRSW}
K.~Alexander, ``The {RSW} theorem for continuum percolation and the {CLT} for
  {E}uclidean minimal spanning trees,'' {\em Ann. Appl. Probab.},  {\bf 6},
  466 (1996).

\bibitem{BCKS}
J.~Chayes, C.~Borgs, H.~Kesten, and J.~Spencer, ``Birth of the infinite
  cluster: finite size scaling in percolation.'' (in preparation).

\bibitem{BS}
I.~Benjamini and O.~Schramm, ``Conformal invariance and {V}oronoi
  percolation.''
\newblock 1996 preprint.

\bibitem{Aiz_prep}
M.~Aizenman.
\newblock In preparation.

\bibitem{KadCeva}
L.~Kadanoff and H.~Ceva, ``Determination of an operator algebra for the
  two-dimensional {I}sing model,'' {\em Phys. Rev. B},  {\bf 3},  3918 (1971).

\bibitem{BTW}
P.~Bak, C.~Tang, and K.~Wiesenfeld, ``Self--organized criticality:  An 
  explanation of $1/f$ noise,'' {\em Phys. Rev. Lett.},  {\bf 59},  381 (1987).

\bibitem{CCN}
J.~Chayes, L.~Chayes, and C.~M. Newman, ``The stochastic geometry of invasion
  percolation,'' {\em Commun. Math. Phys.}, {\bf 101}, 383 (1985).

\bibitem{AlexMST}
K.~Alexander, ``Percolation and minimal spanning forests in infinite graphs,''
  {\em Ann. Probab.},  {\bf 23},  87 (1995).

\bibitem{ABNW} M. Aizenman, A. Burchard, C.M. Newman, and D. Wilson, 
    ``Scaling limits for minimal and random spanning trees in two dimensions,'' 
     in preparation. 
\end{thebibliography}
%redo at the end

\end{document}